\documentclass[aps,
english,preprintnumbers,nofootinbib,
twocolumn]{revtex4-1}

\usepackage{amsfonts,amsmath,amssymb}
\usepackage{graphicx}
\usepackage[utf8]{inputenc}
\usepackage{hyperref}
\usepackage{babel}

\newcommand\be{\begin{equation}}
\newcommand\ee{\end{equation}}
\newcommand\bea{\begin{eqnarray}}
\newcommand\eea{\end{eqnarray}}

\begin{document}

\def\rhoo{\rho_{_0}\!} 
\def\rhooo{\rho_{_{0,0}}\!} 

%


\title{On Quantum Complexity}
\author{Mohsen Alishahiha}
\email{alishah@ipm.ir}
\affiliation{School of Physics, Institute for Research in Fundamental Sciences (IPM),\\
	P.O. Box 19395-5531, Tehran, Iran\\} 


\begin{abstract}
The ETH ansatz for matrix elements of a given operator in the energy eigenstate 
basis results in a notion of thermalization for a chaotic system. 
In this context for a certain quantity - to be found for a given model - one may impose a particular condition 
 on its matrix elements in the energy eigenstate basis so that the corresponding 
 quantity exhibit linear growth at late times. The condition is to do with a possible
 pole structure the corresponding matrix elements may have. Based on the general expectation  of
 complexity one may want to think of this quantity as a possible candidate for 
  the quantum complexity. We note, however, that for the explicit examples we have 
  considered in this paper, there are  infinitely many quantities exhibiting similar behavior.
\end{abstract}


\maketitle

\section{Introduction}
 For chaotic systems with a finite 
entropy $S$, complexity is expected to grow for exponentially large times in the entropy, long after thermal equilibrium has been reached \cite{{Susskind:2014rva},
{Susskind:2014moa}}.
For such systems the notion of thermalization may be described by the
 eigenstate thermalization hypothesis 
(ETH) which gives an understanding of how an observable thermalizes to its thermal
equilibrium value \cite{{Deutsch:1991}, {Srednicki:1994mfb}}(for review
 see \cite{DAlessio:2015qtq}). 

To be concrete let us consider a Hamiltonian, $H$, whose eigenvalues and eigenstates are denoted by $E$ and $|E\rangle$, respectively. Given a general
state $|\psi\rangle$,  the quantum expectation value of an operator, ${\cal O}$,
at given time is 
 \bea\label{QEV}
&&\langle {\cal O}(t)\rangle=\langle \psi|e^{itH}{\cal O}\,e^{-it H}|\psi\rangle\\
&&\;\;\;=
\int_0^\infty dE_1\, dE_2\;
e^{it(E_1-E_2)}\langle \psi|E_1\rangle\langle E_1|
{\cal O}|E_2\rangle\langle E_2|\psi\rangle\nonumber\, .
\eea

In the context of the thermalization of a quantum chaotic system one is typically
 interested in the
equal time averages of observables. More precisely, we would like to find 
 the time average of ${\cal O}$ over a time interval, which will be eventually
 sent to infinity.
 
 According to the ETH, thermalization occurs at the level of
individual eigenstates of the Hamiltonian. In fact setting 
\be\label{Cor}
\varepsilon=\frac{E_1+E_2}{2}, \;\;\;\;\;\;\;\;\omega=E_1-E_2,
\ee 
the
 ETH states that the matrix elements of observables in energy
 eigenstate basis 
 obey the following ansatz \cite{{Srednicki:1999}} 
\be\label{ETH}
\langle E_1|
{\cal O}|E_2\rangle=\bar{\cal O}(\varepsilon) \delta_{E_1,E_2}+e^{-S} f(\varepsilon,
\omega) {\cal R}_{E_1E_2}
\ee
where $\bar{\cal O}(\varepsilon)$ is
the micro canonical average of the corresponding operator, $S$ is thermodynamical 
entropy of the system, $f(\varepsilon,
\omega)$ is a smooth function of its arguments and ${\cal R}$ 
is unit variance random function with zero mean. 

Therefore the quantum expectation value of an observable
satisfying ETH will approach its  thermal equilibrium value given by  the micro canonical average
 for long enough times. 

Of course our main concern in this note is not to explore the thermalization 
process of the system.  Actually the aim of the present letter is to understand the late time behavior 
of a certain observable when the system is in the thermal equilibrium. 

 More precisely, within the context of the ETH, we are interested in
 exploring a possible condition we may impose
on the matrix elements of an observable in energy eigenstate basis so that the corresponding expectation value
exhibits time growth even though the system has been already reached the thermal equilibrium.


\section{Quantum Complexity}

Our main motivation to propose a candidate for the quantum 
complexity comes from the holographic setup in which 
  it is believed 
that the holographic complexity exhibit a linear  
growth at late times \cite{{Susskind:2014rva},
{Susskind:2014moa}}. Therefore, it what follows for a given 
quantum system we would like to 
define a quantity exhibiting such a linear growth.

It is worth noting that the late time linear 
growth is rather a special behavior which is relatively 
well understood in the context of holographic complexity.
In general, it is not clear whether any definition of 
complexity should fulfil such a requirement.  
We note, however, that
for the certain definition of complexity and under 
certain condition 
it is still possible to see the phase of linear growth at 
late times.
We will come back to this point later when we  compare our 
general expression to a particular definition of 
complexity.

 To proceed, following our notation in the previous section 
 let us consider the time dependent expectation value of an 
 operator given by \eqref{QEV}. To be more general, 
 one  may  also  consider the case 
where the system is at finite temperature. 
In this case we will need to consider a time 
shifted state. Thus, let us define the quantum object
${\cal A}$ associated with an operator ${\cal O}$ as 
follows
\be
{\cal A}_{\cal O}(\beta,t)\equiv\langle {\cal O}(t)\rangle_\beta
= \langle \psi|e^{-(\frac{\beta}{2}-it)H}
{\cal O}\,e^{-(\frac{\beta}{2}+it) H}|\psi\rangle\,.
\ee
Using the completeness condition of the energy eigenstates
 \footnote{In 
what follows we will consider a quantum system with 
continuous spectrum. Therefore, our study is 
more appropriate 
for gravity or holographic field theories.}, $\int dE\, |E\rangle\langle E|=1$, one
finds
\bea\label{A}
&&{\cal A}_{\cal O}(\beta,t)=\int_0^\infty dE_1 dE_2\; e^{-\frac{\beta}{2}(E_1+E_2)}
e^{it(E_1-E_2)}\nonumber \\ &&\hspace*{3.3cm}\times\, \rho_\psi(E_1,E_2)
\;A(E_1,E_2),
\eea
where
\bea\label{Af}
\rho_\psi(E_1,E_2)&=&\langle E_1|\psi\rangle\langle\psi|E_2
\rangle=\rho(E_1)\rho(E_2)\,,\cr &&\cr 
A(E_1,E_2)&=&\langle E_1|{\cal O}|E_2\rangle\,,
\eea
with $\rho(E)=\langle E|\psi\rangle$ is the density of state.

As far as the time dependence of the corresponding quantum object is 
concerned, as we will see, the main role 
is played by the function $A$ given in the equation \eqref{Af} 
($A$-function) which is essentially the matrix elements of 
the operator ${\cal O}$ in the energy eigenstates. In particular
one  would expect that for a typical operator the 
$A$-function follows the ETH ansatz \eqref{ETH} and therefore 
the long time average of ${\cal A}_{\cal O}$ approaches  
that of micro canonical average of the corresponding operator.



In our case, however, since we are interested in the late time 
behavior of the quantum object, we will not perform the long time 
average and instead will look for a possible procedure from which the 
late time behavior of ${\cal A}_{\cal O}$ may be read. Actually
we would like to see whether there is a condition under which
the corresponding quantum quantity, ${\cal A}_{\cal O}$, 
keeps growing with time even though the whole system is reached 
thermal equilibrium.

More precisely, in what follows we will explore a possible condition 
one may put on the $A$-function so that the corresponding quantum object ${\cal A}_{\cal O}$ exhibits linear
growth at late times. 
To proceed, since we are interested in the late time behavior, it 
useful to rewrite the expression \eqref{A} in terms of variables 
defined in \eqref{Cor} 
\bea
&&\!\!{\cal A}_{\cal O}(\beta, t)=\int_0^\infty d\varepsilon 
e^{-\beta \varepsilon} \int_{-\infty}^\infty
d\omega\; e^{i\omega t}\rho(\varepsilon+\frac{\omega}{2})\rho(\varepsilon-\frac{\omega}{2})\nonumber \\ & &
\hspace*{5cm}\times\,
A(\varepsilon, \omega),
\eea
and then study the behavior of $A(\varepsilon, \omega)$ in the 
limit of $\omega\rightarrow 0$. 
 
Actually, as it is evident form the above expression, the time 
dependence of ${\cal A}_{\cal O}$ should be read from the 
$\omega$-integral. Indeed, 
due to the simple factor of $e^{i\omega t}$ in the integrand, using 
the Cauchy's residue theorem
\footnote{It is worth noting that the analytic continuation of 
$\omega$ to the complex plane evolves the closing of the 
contour through in the upper half plane where $\omega \rightarrow 
i\infty$. It is then important to make sure that 
the integrand decays along this contour. Due to the presence of the
the exponential factor, $e^{i\omega t}$, the appropriate decay
for whole integrand occurs for a density which 
is bounded in this limit. This should be also 
the case for the finite part of the $A$-function. Although in general,
it is hard to explicitly show the desired behavior for these functions, in 
what follows we will assume that this is the case. Indeed our insight about this assumption
comes for the JT-gravity where analytic form of the integrand is 
known (see the explicit example in the next section) and
it is possible to see, rather explicitly, that 
it has the desired property. We would like to 
thank the referee for his/her comment on this point.} with the assumption 
that the density of 
state $\rho(\varepsilon\pm\omega/2)$ is a smooth function in the limit 
of $\omega\rightarrow 0$, in order to get a non-trivial time 
dependence, the $A$-function must have a pole structure of order of 
$n$ for $n\geq 2$.  
In particular, for the case of a double pole structure where the 
$A$-function has the following limiting behavior
\be\label{a}
A(\varepsilon, \omega)=-\frac{a(\varepsilon)}{\omega^2}+{\rm local\; terms},
\;\;\;\;\;\;
{\rm for}\;\;\omega\rightarrow 0,
\ee
with a positive smooth function $a(\varepsilon)$, one finds that the quantum 
object ${\cal A}_{\cal O}$ exhibits a linear growth at late times
\be\label{LINEAR}
{\cal A}_{\cal O}(\beta,t)=C_0+\int_0^\infty d\varepsilon
e^{-\beta \varepsilon}\rho^2(\varepsilon)a(\varepsilon)\; (2\pi t),
\ee
where $C_0$ is a time independent constant that  is a function of $\beta$.
It is worth noting that for poles of higher order, one generically gets power
 low time dependent behavior. We will come back to this point later.

Motivated by holographic complexity, having found a quantum object exhibiting a linear growth at late times, it is 
tempting to identify the corresponding quantum object, ${\cal A}_{\cal O}$, 
as the quantum complexity. To be precise, we would like to define  complexity 
as follows.

 For a  quantum system the quantum complexity is defined by \eqref{A} 
 for a particular operator
 ${\cal O}$-to be found for a given system- so that the associated $A$-function 
 exhibits a double pole structure in the limit of $E_1\rightarrow E_2$
\be\label{Alate}
A(E_1,E_2)\approx -\frac{a(E_1,E_2)}{(E_1-E_2)^2}+{\rm local\, terms}\,
\ee
where $a(E_1,E_2)$ is a smooth positive function.

Of course for a given quantum system and a given state, a priori, 
 it is not obvious how to find the operator
 ${\cal O}$ that results in the desired 
 double pole structure for $A$-function. 
 Moreover, in general, the corresponding quantity may not be given in terms 
 of local operators. 
 
 It is worth mentioning that there are other quantities which could also  
 exhibit linear growth at late times. These include, for example, the 
 spectrum form factor that has a linear growth known as ramp phase (see for 
 example \cite{Saad:2019pqd}). We note, however, that the linear growth we 
 have found in \eqref{LINEAR} must not be confused with that of the ramp 
 phase mentioned above. Actually,  the ramp phase is a 
 consequence of subleading  connected part of the  density-density 
 correlation, though in our case the linear growth 
at late times occurs at leading disconnected level which is, indeed, a 
distinctive  behavior associated with the complexity. 
 
 Another point we would like to mention is  that the quantum object defined 
 in \eqref{A} is non-local. Our motivation to look for a non-local object as 
 the complexity has mainly come from holography. Indeed the holographic 
 complexity is conjectured to be proportional to the volume of an extremal 
 hypersurface extending all the way behind the 
horizon of a black hole. For two sided eternal black hole the complexity 
is given by the Einstein-Rosen bridge connecting two 
boundaries \cite{Susskind:2014rva,Susskind:2014moa}. In particular in two 
dimensions it is given by the quantum geodesic length connecting two 
boundaries
\cite{Iliesiu:2021ari,Alishahiha:2022kzc,Alishahiha:2022exn}
which is a non-local object. 

Therefore we would expect that quantum complexity should be defined
in terms of a non-local  quantum object. Although in general it might be 
difficult  to understand the natural of non-locality for the quantum object 
\eqref{A}, it is possible to partially understand this for the case where 
our definition of complexity reduces to that of Krylov complexity \cite{AB}.

To further explore our proposal for complexity, in the next section we 
will present an explicit example in which one could identify
 a proper operator
 ${\cal O}$, that results in a linear growth for ${\cal A}_{\cal O}$.

 
\section{Explicit example}

Let us consider a quantum system with the following Hamiltonian
\be\label{LIU}
H=\frac{P^2}{2}+2\mu e^{-x}+2e^{-2x}\,.
\ee
Then the corresponding Schr\"odinger equation is
\be
\left(-\frac{d^2}{dx^2}+4\mu e^{-x}+4e^{-2x}\right)\psi(x)=2E\psi(x)\, .
\ee
 The eigenstate wave functions of the above equation 
are given in terms of the Whittaker function of the second kind with 
imaginary order
\be
\psi_{\mu, E}(x)=e^{x/2}{W_{-\mu, i\sqrt{2E}}(4e^{-x})}\, .
\ee

Actually this Hamiltonian is used to study different aspects of two
dimensional JT gravity (see {\it e.g.} 
\cite{Harlow:2018tqv,Yang:2018gdb,Saad:2019pqd,Gao:2021uro}). For general 
$\mu\neq 0$ it corresponds to JT gravity with end of the world brane whose 
tension is given by $\mu$. 
For the particular value of $\mu=\frac{1}{2}$ 
it may also be considered as supersymmetric version of JT gravity 
\cite{Douglas:2003up}. We note  that this model could be also thought of as a 
Liouville quantum mechanics describing Sachdev-Ye-Kitaev Model 
\cite{Bagrets:2016cdf}.

Using this Hamiltonian the complexity of JT gravity has been also studied in
\cite{Iliesiu:2021ari,Alishahiha:2022kzc,Alishahiha:2022exn}. 
Of course in what follows the relevance 
of this quantum system to the two-dimensional JT gravity is not important for 
us, and we will consider it as a one dimensional quantum system. 

The orthogonality condition for the eigenstates $\psi_{\mu, E}(x)$ is 
\cite{Szmytkowski:2009}
\be
\int_0^\infty dx\;\psi_{\mu, E_1}(x)\;\psi_{\mu,E_2}(x)=
\frac{\delta(E_1-E_2)}{\rho_\mu(E_1)}\,,
\ee
where
\be
\rho_\mu(E)=\left|\Gamma\left(\frac{1}{2}+\mu+i\sqrt{2E}\right)\right|^2 
\frac{\sinh 2\pi \sqrt{2E}}{4\pi^2}\,,
\ee
Following our proposal, the quantum complexity is given by the equation \eqref{A} whose $A$-function,
using the coordinate system, is 
\be
A(E_1,E_2)=\int_0^\infty dx\,dx'\;
 \psi_{\mu, E_1}(x)
 \psi_{\mu, E_2}(x') f(x,x')\, ,
\ee
where $f(x,x')=\langle x|{\cal O}|x'\rangle$. Motivated by the result of \cite{Alishahiha:2022kzc} we will 
consider $f(x, x')=\delta(x-x')\, x$ by which the 
the above $A$-function reads
\be
A(E_1,E_2)=\int_0^\infty dx\;
 \psi_{\mu, E_1}(x)
 \psi_{\mu, E_2}(x) x\, .
\ee
Actually since the function $f$ may be interpreted as 
matrix elements of the operator in the coordinate basis, 
the above choice corresponds to the 
matrix elements of position operator that is
obviously diagonal leading to a delta function. On the other hand since the 
wave function satisfies the Schr\"odinger equation,  in this case, essentially
the $A$-function is 
the average of the position operator. 

By making use of the  explicit expression for the wave function in terms of the Whittaker function, it is then straightforward to study the pole structure of the 
 $A$-function. Indeed, using the variables defined in
\eqref{Cor} and in the limit of 
$E_1\rightarrow E_2$ one finds \cite{Alishahiha:2022kzc}
\be\label{Afun}
A(\varepsilon, \omega)=
-\frac{\sqrt{2\varepsilon}}{2\pi\rho_\mu(\varepsilon)}\;
\frac{1}{\omega^2}+{\rm local\; terms}.
\ee
Therefore from the equation \eqref{A} one can find the late time behavior as
follows 
\bea
{\cal A}(\beta, t)=C_0+\int_0^\infty d\varepsilon\,
e^{-\beta \varepsilon}\rho_\mu(\varepsilon)\sqrt{2\varepsilon}\;t\,
\eea
that is the linear growth, as expected. 

If one recalls that the Hamiltonian \eqref{LIU} describes two 
dimensional JT-gravity it is possible to identify what exactly the 
quantity given in the equation \eqref{A} computes. Indeed in this
case it can be interpreted as the quantum expectation value of the 
geodesic length (wormhole) connecting two boundaries
of a two sided 2d black hole (or a geodesic length connecting the boundary and an end of the world brane) \cite{Alishahiha:2022kzc,Alishahiha:2022exn}. This means 
that the function $f(x,x')$ is just the (regularized) geodesic length. Therefore 
it is expected to get a linear growth behavior at late times.

As another example we note that in the context of random matrix model and its 
connection with chaos we are typically dealing with matter two point 
functions whose matrix elements in energy eigenstate basis have the 
following general form \cite{Jafferis:2022wez}
\be\label{ME}
{\cal O}_{E_1,E_2}=
\frac{|\Gamma(\Delta+i(\sqrt{E_1}-\sqrt{E_2}))\Gamma(\Delta+i(\sqrt{E_1}+
\sqrt{E_2}))|^2}
{\Gamma(2\Delta)}\, .
\ee
where $\Delta$ is the dimension of the corresponding matter field.  
From this expression one may define an $A$-function as follows
\bea
&&A(E_1,E_2)=-\lim_{\Delta\rightarrow 0}\frac{d}{d\Delta}{\cal O}_{E_1,E_2}\\
&&\;\;\;\;\;\;\;\;\;\;\;\;\;\;\;\;\;=-
2|\Gamma(i(\sqrt{E_1}-\sqrt{E_2}))\Gamma(i(\sqrt{E_1}+\sqrt{E_2}))|^2.
\nonumber
\eea
It is then easy to see that in the limit of $E_1\rightarrow E_2$ this 
$A$-function exhibits a double pole structure 
\be
A(E_1,E_2)=-\frac{4\pi \sqrt{\varepsilon}}{\sinh(2\pi \sqrt{\varepsilon})
\omega^2}+{\rm local\, terms}\, .
\ee
In fact, recalling the relation between random matrix model and 
two-dimensional JT gravity, the above expression 
corresponds to the case of $\mu=0$ in \eqref{Afun}.

It is also interesting to look at the rate of the complexity growth
\be
\frac{d{\cal A}_f(t)}{dt}=\int_0^\infty d\varepsilon
e^{-\beta \varepsilon}\rho_\mu(\varepsilon)\sqrt{2\varepsilon}\,,
\ee
which may be compared with the Lloyd's bound \cite{Lloyd:2000}. Actually,
 in the context of holographic complexity in which the complexity may
be computed using CA conjecture \cite{Brown:2015bva} the rate of the 
complexity growth turns out to be twice of the energy of the system, 
saturating the Lloyd's bound \cite{Lloyd:2000}. In the present case, 
at low energies where  $\rho_\mu(\varepsilon)\sim \sqrt{2\varepsilon}$  
the above expression may be thought of as the average of energy in 
a canonical ensemble.

On the other hand, if one works with a non-normalized situation by
 dropping $1/Z$ factor,
 one could evalaute  the rate of the complexity growth in the macro canonical
 ensemble by making use of  the inverse  Laplace transformation. In this case 
 the corresponding
rate is given by $\rho(E_0) \sqrt{2E_0}$ which at low energies results in
$\sim 2E_0$, reminiscing of the Lloyld's bound. Here $E_0$ is the energy 
of the macro canonical ensemble.

In general  for $\mu=0$ the integral may be performed exactly to find
\be
\frac{d{\cal A}_f(t)}{dt}=\frac{2e^{-\frac{2\pi^2}{\beta}}}{\sqrt{2\pi 
\beta}}+
\left(\frac{1}{2\pi}+\frac{4\pi^2}{\beta}\right){\rm Erf}\left(\frac{\sqrt{2}
\pi}{\sqrt{\beta}}\right),
\ee
which at low temperatures goes as $\sim\beta^{-1/2}$ while at high 
temperatures it
is $\sim\beta^{-1}$. Although for general $\mu$ the full expression for
the rate of the complexity growth may not be written explicitly, asymptotic 
behaviors at low and high temperatures are the same as that of $\mu=0$.

 
\section{Discussions}

In this letter we have defined a quantum object associated with a given 
operator in a chaotic system. We have demonstrated that under certain 
condition the corresponding quantum object exhibits linear growth at 
late times, much longer than the system reaches the thermal equilibrium. 

We have shown that for
a given operator ${\cal O}$-to be found for given system- if its
 matrix elements in the energy eigenstates exhibit 
a double pole structure at late times \eqref{Alate}, the corresponding 
quantum object defined it the equation \eqref{A} will have the linear 
growth at late times, which could be interpreted as the quantum complexity.
Of course for a given state in a given system, a priori, it is not clear how 
to find the operator  ${\cal O}$ with the above desired property. It is not 
even clear if it is a local operator.

In the context of thermalization of quantum system, one 
generally assumes that matrix elements follow the ETH 
ansatz. Though, as we have seen, in order to get a non-trivial
time dependence at late times, the $A$-function should have poles of order 
of $n$ with $n\geq 2$. For $n>2$ one generally gets power low
growth at late times. For $n=2$ it is a linear growth. 
Since having a linear growth  at late times (at leading order) is a 
signature of the complexity \cite{Susskind:2014rva} which 
 is expected to be the fastest growth, one may propose a hypothesis 
 that the double pole structure is the highest pole structure the 
 $A$-function could have.

It is worth mentioning that for a give chaotic model there could be several
$f$'s (matrix elements in coordinate basis) that give double pole structure for $A$-function which  result in the
late time linear growth. For the explicit example we have presented in the
previous section it is straightforward to see that for any functions in 
the form of $f(x,x')=\delta(x-x') x^m$, with integer $m$, one finds 
double pole structure. Moreover form the matrix elements \eqref{ME} it 
is easy to construct several $A$-functions with the desired property. 
They can be obtained by taking $\Delta\rightarrow
0$ limit of $m^{th}$ $\Delta$-derivative of the matrix elements \eqref{ME}.

This is very similar to the observation made in \cite{Belin:2021bga} in 
the context of the holographic complexity where it was shown that there 
are infinite class of gravitational observables in asymptotically 
Anti-de Sitter space which living on codimension one slices of the 
geometry, that exhibit universal features as that in complexity. 
Namely they grow linearly in time at late times.


An other interesting feature of complexity is that it saturates at the very 
late times given by the exponential of the entropy of the system. It is then 
natural to see how the saturation could occur in this context. 

To address this question we note that the
the density matrix $\rho(E_1,E_2)$ appearing in the 
expression of the quantum object ${\cal A}_{\cal O}$ \eqref{A} 
has the following general form
\be\label{rho}
 \rho(E_1,E_2)=\rho(E_1)\rho(E_2)+\rho_{\rm c}(E_1,E_2)\, .
 \ee
where $\rho_{\rm c}$ represents the connected term meaning that it cannot be 
written in a factorized form of $g_1(E_1)g_2(E_2)$ with $g_{1,2}$ being 
arbitrary functions.
Clearly form the first factorized term the above function reduces to that of 
\eqref{A}.
The connected terms could have either perturbative or non-perturbative 
origins which may have  generally non-trivial pole structure that could 
result in the saturation phase at very late times.

Actually this is a well known structure which has been seen in the literature 
for the spectrum form factor of chaotic models such as JT-gravity in
which the pole structure of $\rho(E_1,E_2)$ results in the ramp phase. 
Of course for the spectrum form factor there is no an $A$-function
and the whole time dependence is controlled  by the  density-density 
correlator. On the other hand, for the holographic complexity of JT-gravity
where there is  an $A$-function, the connected 
part of $\rho(E_1,E_2)$, which has non-trivial pole structure at late
times is, indeed, responsible for the saturation phase 
\cite{Iliesiu:2021ari,Alishahiha:2022kzc,Alishahiha:2022exn}.

We note, however, that in the present case, where we are dealing with a
general formalism which is not directly related to the holography picture,
it is not clear how the full expression of the connected term could be 
computed. Nonetheless, for a chaotic system,  as far as the late time 
behavior is concerned, one would expect
that the main contribution comes from the short range correlation which
is given by the universal sine-kernel term \cite{Mehta:2004}\footnote{ I 
would like to thank Julian Sonner for pointing out  this to me.}
\be\label{CON}
\rho_c(\varepsilon,\omega)\approx -\frac{\sin^2(D\omega \rho(\varepsilon))}
{(D\omega)^2},\;\;\;\;\;{\rm for}\;\omega\ll 1\,.
\ee
 Here $D$ is the dimension of Hilbert space which
is given by the exponential of the entropy of the system. Therefore the whole
late time behavior of the quantum object ${\cal A}$ is described as 
follows: the double point structure of the $A$-function leads to linear 
growth at the leading disconnected part of the density-density correlation, 
while  there is the saturation phase which can be described by subleading 
connected term given in the universal sine-kernel term multiplied by the 
double pole structure of the $A$-function. It is then easy to see
that the saturation occurs at $t\sim D$.

To be precise plugging the expression \eqref{rho} into \eqref{A} and using 
\eqref{CON} and \eqref{a} one arrives at
\bea
\label{eq:lt}
{\cal A} &=&{\rm Constant}-
\int_0^\infty d\epsilon e^{-\beta \epsilon} \rho^2(\epsilon)a(\epsilon)
\\ &&\int_{-\infty}^\infty d\omega 
\frac{e^{-it\omega}}{\omega^2}\left(1-\frac{\sin^2\left({\rho}(\epsilon) D\omega\right)}{(\rho(\epsilon) D \omega)^2}\right)\,.\nonumber\\&&\nonumber
\eea

From this expression one observes that  at late times 
when $\omega\sim \frac{1}{t}\rightarrow 0$ and 
for $\rho \, \omega\gg 1$ essentially the first term in the bracket on the r.h.s of  \eqref{eq:lt} dominates leading 
to a linear growth, while for $\rho \, \omega\ll  1$ which is the case at $t \sim D$, the second term starts dominating that essentially cause the whole integral to approach zero
leading to a constant complexity which is the saturation phase.
For more details see \cite{Iliesiu:2021ari,Alishahiha:2022kzc,Alishahiha:2022exn}.

It is worth noting that due to the particular form of the quantum object ${\cal A}_{\cal O}$ \eqref{A} in which the $A$-function and $\rho(E_1,E_2)$ appear in a product form, there is an alternative way to think about the saturation phase.
Indeed the saturation may occur via an ETH-like behavior of 
$A$-function at very late times. In this case the disconnected part of density matrix 
is enough to see the saturation phase.

As a final comment we note that the structure we have presented in this 
letter has similar features as that of the Krylov complexity 
\cite{Parker:2018yvk,Barbon:2019wsy,Jian:2020qpp,Rabinovici:2020ryf,Kar:2021nbm,
Balasubramanian:2022tpr,Balasubramanian:2022dnj}. 
Actually, it can be shown that for a particular case our proposal for complexity  reduces to that of Krylov complexity. In this case  the $A$-function is given by 
the matrix elements in energy basis of a {\it label operator} of an 
orthonormal and ordered basis (Krylov basis). More precisely, denoting the orthonormal ordered basis by $\{|n\rangle\}$, the label operator is defined by $\ell=\sum_{n}n\,|n\rangle\langle n|$. Therefore one gets ${\cal A}_f(0,t)=\langle \ell(t)\rangle$ with $\ell(t)=e^{-iHt}\,\ell\, e^{iHt}$ and the $A$-function reads
$A=
\langle E_1|\ell|E_2\rangle$. Although the linear growth in the Krylov complexity is rather a special case which may occur when the Lanczos coefficients saturate to a constant\footnote{
We note that the late time linear growth in the context 
of Krylov complexity for generic chaotic systems has been first observed in \cite{Barbon:2019wsy} where it was shown that 
the saturation of Lanczos coefficients to a constant results in 
the linear growth at sufficiently late times, much after 
the scrambling time.},  it is  possible to perform explicit 
computations (at least in the continuum limit) to show that the saturation
of Lanczos coefficients corresponds to the double pole structure of the $A$-function at late times (for more details see \cite{AB}).
Therefore our proposal would appropriately reproduce the linear growth of the Krylov complexity known in the literature \cite{Barbon:2019wsy}. Moreover, in this case the 
saturation occurs due to an ETH-like behavior of the label operator.

 \begin{acknowledgments}
I would like to think Alexandre Belin, Kyriakos Papadodimas and Julian Sonner for discussions. The author would also like to thank Sonner's group in Geneva University
for discussions and comments.
I would also like to thank  Souvik Banerjee for comments  and 
 discussions on different aspects of complexity as well as collaborating on 
 several past and ongoing projects. 
The author would also like to thank Department of Theoretical Physics of CERN for very warm hospitality

\end{acknowledgments}

\bibliographystyle{apsrev4-1}


\begin{thebibliography}{}

\bibitem{Susskind:2014rva}
L.~Susskind,
``Computational Complexity and Black Hole Horizons,''
Fortsch. Phys. \textbf{64} (2016), 24-43
doi:10.1002/prop.201500092
[arXiv:1403.5695 [hep-th]].



\bibitem{Susskind:2014moa}
L.~Susskind,
``Entanglement is not enough,''
Fortsch. Phys. \textbf{64} (2016), 49-71
doi:10.1002/prop.201500095
[arXiv:1411.0690 [hep-th]].

\bibitem{Deutsch:1991}
J.~M.~Deutsch, ``Quantum statistical mechanics in a closed system,''
Phys Rev A.{\bf 43} 2046 (1991),
 doi :10.1103/PhysRevA.43.2046.
 

\bibitem{Srednicki:1994mfb}
M.~Srednicki,
``Chaos and Quantum Thermalization,''
doi:10.1103/PhysRevE.50.888
[arXiv:cond-mat/9403051 [cond-mat]].



\bibitem{DAlessio:2015qtq}
L.~D'Alessio, Y.~Kafri, A.~Polkovnikov and M.~Rigol,
``From quantum chaos and eigenstate thermalization to statistical mechanics and thermodynamics,''
Adv. Phys. \textbf{65} (2016) no.3, 239-362
doi:10.1080/00018732.2016.1198134
[arXiv:1509.06411 [cond-mat.stat-mech]].


\bibitem{Srednicki:1999}
M.~Srednicki, ``The approach to thermal equilibrium in quantized chaotic systems,''
J. Phys.  A{\bf 32} (1999) 1163,
doi:10.1088/0305-4470/32/7/007.
 
\bibitem{Saad:2019pqd}
P.~Saad,
``Late Time Correlation Functions, Baby Universes, and ETH in JT Gravity,''
[arXiv:1910.10311 [hep-th]].
 
 
\bibitem{AB}
M.~Alishahiha and S.~Banerjee,
``A universal approach to Krylov State and Operator complexities,''
[arXiv:2212.10583 [hep-th]].
 

\bibitem{Harlow:2018tqv}
D.~Harlow and D.~Jafferis,
``The Factorization Problem in Jackiw-Teitelboim Gravity,''
JHEP \textbf{02} (2020), 177
doi:10.1007/JHEP02(2020)177
[arXiv:1804.01081 [hep-th]].


\bibitem{Yang:2018gdb}
Z.~Yang,
``The Quantum Gravity Dynamics of Near Extremal Black Holes,''
JHEP \textbf{05} (2019), 205
doi:10.1007/JHEP05(2019)205
[arXiv:1809.08647 [hep-th]].



\bibitem{Gao:2021uro}
P.~Gao, D.~L.~Jafferis and D.~K.~Kolchmeyer,
``An effective matrix model for dynamical end of the world branes in Jackiw-Teitelboim gravity,''
JHEP \textbf{01} (2022), 038
doi:10.1007/JHEP01(2022)038
[arXiv:2104.01184 [hep-th]].


\bibitem{Douglas:2003up}
M.~R.~Douglas, I.~R.~Klebanov, D.~Kutasov, J.~M.~Maldacena, E.~J.~Martinec and N.~Seiberg,
``A New hat for the c=1 matrix model,''
[arXiv:hep-th/0307195 [hep-th]].


\bibitem{Bagrets:2016cdf}
D.~Bagrets, A.~Altland and A.~Kamenev,
``Sachdev\textendash{}Ye\textendash{}Kitaev model as Liouville quantum mechanics,''
Nucl. Phys. B \textbf{911} (2016), 191-205
doi:10.1016/j.nuclphysb.2016.08.002
[arXiv:1607.00694 [cond-mat.str-el]].





\bibitem{Iliesiu:2021ari}
L.~V.~Iliesiu, M.~Mezei and G.~S\'arosi,
``The volume of the black hole interior at late times,''
JHEP \textbf{07}, 073 (2022)
doi:10.1007/JHEP07(2022)073
[arXiv:2107.06286 [hep-th]].

\bibitem{Alishahiha:2022kzc}
M.~Alishahiha, S.~Banerjee and J.~Kames-King,
``Complexity via replica trick,''
JHEP \textbf{08} (2022), 181
doi:10.1007/JHEP08(2022)181
[arXiv:2205.01150 [hep-th]].

\bibitem{Alishahiha:2022exn}
M.~Alishahiha and S.~Banerjee,
``On the saturation of late-time growth of complexity in supersymmetric JT gravity,''
[arXiv:2209.02441 [hep-th]].




\bibitem{Szmytkowski:2009},
 R.~Szmytkowski and S.~Bielski, ``An orthogonality relation for the Whittaker functions of the second kind of imaginary order,'' 
arXiv:0910.1492 [math.CA]





\bibitem{Jafferis:2022wez}
D.~L.~Jafferis, D.~K.~Kolchmeyer, B.~Mukhametzhanov and J.~Sonner,
``JT gravity with matter, generalized ETH, and Random Matrices,''
[arXiv:2209.02131 [hep-th]].








\bibitem{Lloyd:2000}
S.~Lloyd, ``Ultimate Physical limits to computation,''
Nature {\bf 406} (2000) 1047, quant-ph/9908043.


\bibitem{Brown:2015bva}
A.~R.~Brown, D.~A.~Roberts, L.~Susskind, B.~Swingle and Y.~Zhao,
``Holographic Complexity Equals Bulk Action?,''
Phys. Rev. Lett. \textbf{116} (2016) no.19, 191301
doi:10.1103/PhysRevLett.116.191301
[arXiv:1509.07876 [hep-th]].




\bibitem{Belin:2021bga}
A.~Belin, R.~C.~Myers, S.~M.~Ruan, G.~S\'arosi and A.~J.~Speranza,
``Does Complexity Equal Anything?,''
Phys. Rev. Lett. \textbf{128} (2022) no.8, 081602
doi:10.1103/PhysRevLett.128.081602
[arXiv:2111.02429 [hep-th]].



\bibitem{Mehta:2004}
M. L. Mehta,`` Random Matrices.'' (Elsevier, San Diego, 2004) Third edition






\bibitem{Parker:2018yvk}
D.~E.~Parker, X.~Cao, A.~Avdoshkin, T.~Scaffidi and E.~Altman,
``A Universal Operator Growth Hypothesis,''
Phys. Rev. X \textbf{9} (2019) no.4, 041017
doi:10.1103/PhysRevX.9.041017
[arXiv:1812.08657 [cond-mat.stat-mech]].

\bibitem{Barbon:2019wsy}
J.~L.~F.~Barb\'on, E.~Rabinovici, R.~Shir and R.~Sinha,
``On The Evolution Of Operator Complexity Beyond Scrambling,''
JHEP \textbf{10} (2019), 264
doi:10.1007/JHEP10(2019)264
[arXiv:1907.05393 [hep-th]].

\bibitem{Jian:2020qpp}
S.~K.~Jian, B.~Swingle and Z.~Y.~Xian,
``Complexity growth of operators in the SYK model and in JT gravity,''
JHEP \textbf{03} (2021), 014
doi:10.1007/JHEP03(2021)014
[arXiv:2008.12274 [hep-th]].


\bibitem{Rabinovici:2020ryf}
E.~Rabinovici, A.~S\'anchez-Garrido, R.~Shir and J.~Sonner,
``Operator complexity: a journey to the edge of Krylov space,''
JHEP \textbf{06} (2021), 062
doi:10.1007/JHEP06(2021)062
[arXiv:2009.01862 [hep-th]].

\bibitem{Kar:2021nbm}
A.~Kar, L.~Lamprou, M.~Rozali and J.~Sully,
``Random matrix theory for complexity growth and black hole interiors,''
JHEP \textbf{01} (2022), 016
doi:10.1007/JHEP01(2022)016
[arXiv:2106.02046 [hep-th]].


\bibitem{Balasubramanian:2022tpr}
V.~Balasubramanian, P.~Caputa, J.~M.~Magan and Q.~Wu,
``Quantum chaos and the complexity of spread of states,''
Phys. Rev. D \textbf{106} (2022) no.4, 046007
doi:10.1103/PhysRevD.106.046007
[arXiv:2202.06957 [hep-th]].

\bibitem{Balasubramanian:2022dnj}
V.~Balasubramanian, J.~M.~Magan and Q.~Wu,
``A Tale of Two Hungarians: Tridiagonalizing Random Matrices,''
[arXiv:2208.08452 [hep-th]].






\end{thebibliography}

\end{document}